\documentclass[preprint2,times,tighten]{aastex61}
\usepackage{graphicx,graphics,amsmath}
\usepackage{natbib}
\usepackage{bm,url}
\usepackage{times}
\usepackage{amssymb}
\usepackage{color}
\usepackage{float}
\usepackage{arydshln}
\graphicspath{{./fig/}{./png/}}
\usepackage{color}

\def\bl{Babcock--Leighton}
\newcommand{\Fig}[1]{Figure~\ref{#1}}
\newcommand{\Eq}[1]{Equation~(\ref{#1})}
\newcommand{\Sec}[1]{Section~\ref{#1}}

\newcommand{\mps}{m~s$^{-1}$}
\newcommand{\cmss}{cm$^2$~s$^{-1}$}

\shorttitle{Cause of double-peaks in the solar cycle}
\shortauthors{Karak, Mandal \& Banerjee}
\begin{document}

\title{Double-peaks of the solar cycle: An explanation from a dynamo model}

\correspondingauthor{Bidya Binay Karak}
\email{karak.phy@iitbhu.ac.in}
\author{Bidya Binay Karak}
\affil{Department of Physics, Indian Institute of Technology (Banaras Hindu University), Varanasi, India}
\author{Sudip Mandal} 
\affil{Indian Institute of Astrophysics, Koramangala, Bangalore 560034, India}
\author{Dipankar Banerjee}
\affil{Indian Institute of Astrophysics, Koramangala, Bangalore 560034, India}
\affil{Center of Excellence in Space Sciences India, IISER Kolkata, Mohanpur 741246, West Bengal, India}

\begin{abstract}
One peculiar feature of the solar cycle which is yet to be understood properly is the
frequent occurrence of double peaks (also known as the Gnevyshev peaks).
Not only the double peaks but also multiple peaks and spikes are often observed in any phase
of the cycle.
We propose that these peaks and spikes are generated due to fluctuations in the \bl\ process (the poloidal field generation from tilted bipolar magnetic regions). 
When the polar field develops, large negative fluctuations in the \bl\ process can reduce the net polar field abruptly. As these fluctuations in the polar field are propagated to the new toroidal field, these can promote double peaks in the next solar cycle. When fluctuations in the polar field occur outside the solar maximum, we observe their effects as spikes or dips in the following sunspot cycle.      
Using an axisymmetric \bl\ dynamo model we first demonstrate this idea. Later, we perform a long simulation by including random scatter in the poloidal field generation process and successfully reproduce the double-peaked solar cycles.
These results are robust under reasonable changes in the model parameters 
as long as the diffusivity is not too larger than $10^{12}$~cm$^2$~s$^{-1}$.
Finally, we analyze the observed polar field data to show a close connection between the short-term fluctuations in the polar field and the double peaks/spikes in the next cycle.
Thereby, this supports our theoretical idea that the fluctuations in the \bl\ process 
can be
responsible for the double peaks/spikes in the observed solar cycle.
\end{abstract}



\section{Introduction}
\label{sec:int}
Sun's magnetic activity, commonly measured using the sunspot number or sunspot area, 
oscillates with a period of about 11 years. This is
popularly known as the solar cycle or sunspot cycle. 
Interestingly, every solar cycle is different from the previous ones 
in terms of the cycle duration and amplitude.
Apart from this variation, there exist several short-term variations 
in the observed solar data \citep{LB89,Baz14,Scott15,Maetal}.

One distinct and puzzling observable among these short-term variations is
the occurrences of double peaks. It has been observed that during the solar maximum, 
when sunspot number reaches its maximum value,
solar cycle occasionally shows two peaks \citep{FS97,NG10,Geor,Baz14}.
These are also known as {\it Gnevyshev peaks} and the 
gap between these two peaks at the solar maximum is known 
as {\it Gnevyshev gap} \citep{Gnev67,Gnev77}.
Although observed in many earlier cycles, this double-peak feature has received 
special attention in recent years mainly because of the last three solar cycles 
being double-peaked; see NASA science 
report\footnote{\href{https://science.nasa.gov/science-news/science-at-nasa/2013/01mar_twinpeaks}{https://science.nasa.gov/science-news/science-at-nasa/2013/01mar$\_$twinpeaks}}.
We note that these double peaks  
are not the artefacts of insufficient observations 
but are real features \citep{NG10}.
We also note that this feature is not only limited to the sunspot number or area data,
but also observed in other proxies of the solar activity e.g., 
coronal activity \citep{Gnev63,Kane09, Kane10}

One could argue that the double peak is a result of the fact that when two hemispheres
reach their maxima
at two different times, the combined solar activity can have two peaks.
By making a careful analysis of the solar data, we shall show that a time difference
between the maxima of two hemispheric solar activity may lead to a double peak,
however, this happens rarely. In fact, most of the times, the double-peak occurs only in one hemisphere \citep{NG10}.
Importantly, the double-peak type spikes are not only observed during solar maximum, they are
also seen at any phase of the solar cycle. When the spike appears near a solar maximum, we see
it as a double-peak. 

The double peaks and spikes are possibly the manifestrations of the recently discovered 
quasiperiodic ``burst" or oscillations with periods of 6 -- 18 months in the solar activity \citep{Scott15}.
Using magnetohydrodynamics shallow-water model, \citet{Dik17,Dik18} have shown that the
energy exchange among magnetic fields, Rossby waves and
differential rotation in the solar tachocline can lead to quasi-peridic nonlinear oscillations, 
which possibly correspond to the observed burst of solar activity. 
Also see \citet{Za10,Za18} for studies connecting the Rossby waves in the tachocline 
with the short-term oscillations.

However, there could be a different mechanism of producing double peaks and spikes in the solar cycle.
Irregular fluctuations are inherent 
in the solar dynamo which can appear in any phases of the solar cycle. When strong fluctuations
appear near the solar maximum, we may see them as double peaks.
In this study, using a dynamo theory we shall identify the source of these fluctuations and explore how these
fluctuations can promote double peaks in the solar cycle.

Over last two decades, the solar magnetic cycle has been modelled with great details
using the \bl\ dynamo models, also named as the flux transport dynamo models 
\citep{CSD95,Dur95,DG09,Cha10,Kar14a}. 
In this model,
the poloidal field is generated from the decay and dispersal of tilted bipolar magnetic regions (BMRs)
near the solar surface. This field is largely transported to the poles through meridional flow. 
From the surface, the poloidal field is then transported down to the deep convection zone (CZ)
through meridional circulation, turbulent diffusion and pumping, where differential rotation
stretches this field to produce a toroidal field. The toroidal field then rises up the surface
due to magnetic buoyancy and gives tilted BMR. It is believed that the tilt is introduced
due to the Coriolis force during the rise of the toroidal flux in the CZ \citep{DC93}.
The observed correlation between the surface polar flux and the next cycle strength supports this part of the dynamo model \citep{Das10,KO11,Muno13,Priy14}.
The new BMRs again decay and produce poloidal flux which forms the seed for the next cycle.

The tilt angle of a BMR is crucial in generating a net poloidal flux as has been realized in the surface observations \citep{Das10},
as well as surface flux transport models \citep{JCS14}, and a 3D (or 2$\times$2D coupled) dynamo model with explicit BMR depositions \citep{HCM17, KM17, LC17}. In observations, 
we find a considerable scatter of the mean BMR tilt around its systematic variation 
with the latitude---Joy's law \citep{How91,SK12,MNL14,pavai15,MN16}. 
This scatter is the primary cause of the variation in the polar field \citep[e.g.,][]{JCS14,HCM17,KM17, Nagy17}. 
The effect of scatter is very profound when BMRs appear near the equator \citep{Ca13,KM18}.
Other effects such as the fluctuations in the net BMR flux, BMR emergence rates, 
time delay of BMR emergence, meridional circulation speed etc 
can also introduce additional variation in the polar flux 
\citep{LC17,KM17,Nagy17}.
Ultimately, it is the fluctuations in the \bl\ process which is the primary 
cause of the variation in the polar field and 
consequently in the sunspot cycle as has been pointed out earlier by 
\cite{CD00, CCJ07, CK09}.

In this study, we shall show that the fluctuations in the \bl\ process can also 
occasionally produce short-term fluctuations in the polar field. 
These fluctuations can be propagated to the toroidal field and therefore can cause double peaks in the next solar cycle. As these fluctuations can occur at any phase of the polar field build up, the fluctuations can appear at any phase of the solar cycle. 
When they occur outside the solar maxima, we observe them as spikes and dips.
We shall explicitly identify the fluctuations in the polar field from the observed data and show
that this 
can be responsible for 
the double peaks in the solar cycle.

\begin{figure*}
\centering
\includegraphics[scale=0.70]{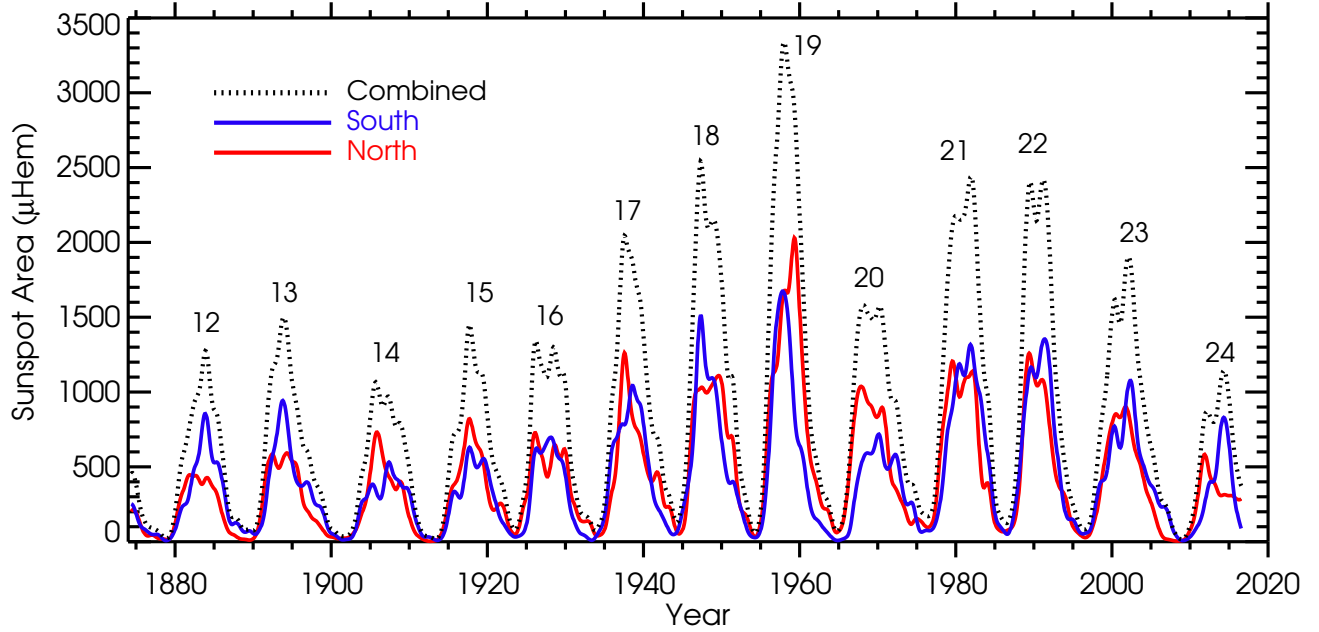}
\caption{Temporal variation of the monthly sunspot area (in millionths of a solar hemisphere) 
for individual hemispheres as well as the combined data. 
Cycle numbers are also printed on the figure.
}
\label{fig:obsssn}
\end{figure*}


\section{Confirmation of double peaks in the solar data}
\label{sec:obs}
Before we discuss the physical source of the double-peak, let us first re-establish its 
existence in the observed solar data. For this purpose, we utilize the Greenwich/NOAA sunspot area 
data\footnote{\url{https://solarscience.msfc.nasa.gov/greenwch.shtml}} 
which covers a period of $\sim$140 years. 
We use sunspot area data instead of sunspot number because a longer hemispheric sunspot number 
is not available.
In order to bring out the prominent spikes in the solar cycle, 
we smooth the monthly averaged sunspot area data with 
a Gaussian smoothing filer of~$\mathrm{FWHM} = 1$ year.  
\Fig{fig:obsssn} shows these smoothed data for the northern (red curve), 
southern (blue) and the combined (dotted) hemispheres. 
When we look at the combined data, we observe that cycles 14, 16, 20, 21, 22, and 23 
have definite double peaks.
However, upon examining the individual hemispheric data, we find
that many cycles have double peaks and spikes only in one hemisphere.
In fact, almost all cycles, except cycles 17 and 19, have double peaks or even multiple peaks. 
But only for a few cycles (16, 20, 21, and 22), double peaks occurred in both hemispheres.
We also note that the peaks are not limited to the solar maxima, they
are seen in the rising or declining phase of the cycle as well; 
see the northern hemisphere of cycles: 17 and 21 and both hemispheres of cycles: 15, 18, and 20.

As discussed in the Introduction, the double peak might appear
when the peaks of two hemispheric activities do not synchronize. This has happened for cycles 22 
and 24 in which north and south hemispheres do not peak at the same time 
and the net sunspot area becomes double-peaked.
However, this cannot happen always. 
For example, cycles 12, 14, 17, 18, and 19 have little time lags between two hemispheric maxima but do not show clear double peaks.
Therefore, we believe that the double-peak is real. This is also in agreement with the analysis of \citet{NG10}.  Moreover,
occasionally we observe multiple peaks (cycles: 16 and 20).
Therefore, this supports our initial guess about the fluctuations in the solar dynamo process 
that is responsible for producing these spikes and double peaks.


\section{Theoretical Model}
\label{sec:mod}
In this study, we use a kinematic axisymmetric \bl\ dynamo model in which
we solve following equations in the solar convection zone \citep{CNC04}.
\begin{equation}
 \frac{\partial A}{\partial t} + \frac{1}{s}({\bf v}\cdot{\bf \nabla})(s A)
 = \eta_{p} \left( \nabla^2 - \frac{1}{s^2} \right) A + \alpha B,~~~~~~~~~~~~~~~~~~~~~~~~~~~~~~~~~~~~~~~~
\label{eq:pol}
\end{equation}
\begin{eqnarray}
 \frac{\partial B}{\partial t} 
 + \frac{1}{r} \left[ \frac{\partial}{\partial r}
 (r v_r B) + \frac{\partial}{\partial \theta}(v_{\theta} B) \right]
 = \eta_{t} \left( \nabla^2 - \frac{1}{s^2} \right)B ~~~~~~~~~~~~~~ \nonumber\\
 + s({\bf B}_p.\nabla)\Omega
 + \frac{1}{r}\frac{d\eta_t}{dr}\frac{\partial{B}}{\partial{r}},~~~~~~~~~~~~~~
\label{eq:tor}
\end{eqnarray}
where $A$ and $B$ are the potential of the poloidal magnetic field ($\bf {B}_p$) and 
the toroidal magnetic field, respectively such that
${\bf B} =  {\bf B}_p + B {\bf \hat{e_{\phi}}}$, 
with ${\bf B}_p = {\bf \nabla} \times A {\bf \hat{e_{\phi}}}$, 
$s = r \sin \theta$ with $\theta$ being the colatitude, 
${\bf v}=v_r {\bf \hat{e_r}} + v_{\theta} {\bf \hat{e_{\theta}}}$ 
is the meridional flow, $\Omega$ is the angular velocity, $\eta_p$ and $\eta_t$ are the diffusivities of the
poloidal and toroidal fields, respectively, $\alpha$ is the coefficient 
describing the generation of the poloidal field from the toroidal field 
and mimics the \bl\ process. Thus
\begin{eqnarray}
\alpha = \frac{\alpha_0}{4}\cos\theta \left[1 + \mathrm{erf} \left(\frac{r - 0.95R}{0.025R}\right) \right]\nonumber \\
\times \left[1 - \mathrm{erf} \left(\frac{r - R}{0.025R}\right) \right],
\label{eq:alpha}
\end{eqnarray}
where $\alpha_0=50~$\mps and $R$ is the solar radius.

We shall not describe details of other ingredients of the model here but refer the readers to
the key publication by \citet{CNC04}. The exact parameters used in this publication 
are the same as given in \citet{YNM08} and \citet{KN12} with the key parameters:
$v_0 =$ 26~\mps, $\eta_2 = 1\times10^{12}$~\cmss, and $\eta_0 = 2\times10^{12}$~\cmss.

Our \bl\ type dynamo model (including many recent models, e.g., \citet{MD14, LC17}) 
although produces many basic features of the solar cycle reasonably well, 
it produces much stronger polar field at the surface than the 
present observational reported values. 
\citet{DG01,Dik02} have shown that this problem can be elevated (at least partially) 
by increasing the surface diffusivity 
of the magnetic field and adding an additional source for the poloidal field in the tachocline.
On the other hand, \citet{KN16} have shown
that a diamagnetic pumping near the base of the convection zone can help.
However, as many of the parameters in this model are not constrained
by observations and no high-resolution magnetograms of the polar magnetic field are available at present,
we ignore the discrepancy between the theory and observations in this work; also see the discussion in \citet{Ch03}.

\section{Theoretical Results}
\label{sec:res}
\subsection{Demonstrating the idea}
\label{sec:idea}
\begin{figure}
\centering
\includegraphics[scale=0.70]{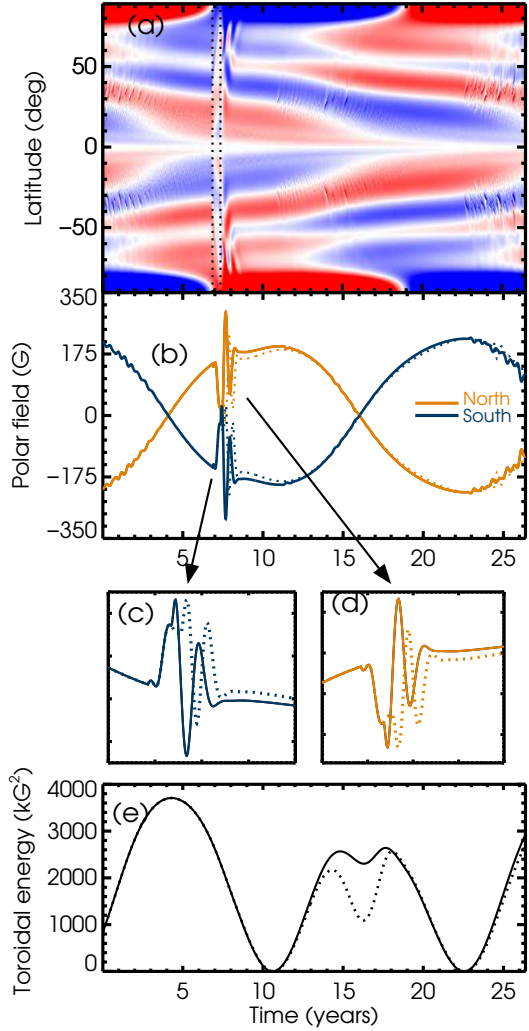}
\caption{
(a): Butterfly diagram of the surface radial field. Two dotted vertical lines show the 6-month
window during which the $\alpha_0$ was reversed.
(b): The mean surface polar field (in G) computed within 
latitudes of $\pm 55^\circ$ -- $\pm 89^\circ$.
(c) and (d): Same as (a) but enlarged view of the polar fields 
shown for 5--10 years.
(e) Toroidal magnetic energy in kG$^2$ obtained 
at $r=0.7R$ and $15^\circ$ latitude.
Dotted lines in (b)--(e) are obtained from a different simulation 
in which the $\alpha_0$ was reversed for 8 months instead of 6 months.
}
\label{fig:testidea}
\end{figure}

Before we present our theoretical results of the double-peaked solar
cycle, let us first describe our idea. We propose that the double peaks (including
multiple peaks and spikes) in the solar cycle are caused by the fluctuations
in the \bl\ process of generating the poloidal field. To demonstrate that this idea is working
in our model in the first place, we do the following experiment. 
We take our dynamo model as described above and when it is producing 
a stable/relaxed solution, we stop the model
at a solar maximum when the polar field has started developing (t = 6.85~year in \Fig{fig:testidea}. 
Then we reverse the $\alpha$ i.e., we make $\alpha_0 = -$50~\mps\ in \Eq{eq:alpha}
in both hemispheres and continue the run for 6~months. After that, we change $\alpha_0$ back to 
50~\mps\ and extend the run for some years.

As soon as $\alpha_0$ gets flipped, the model generates an opposite polarity poloidal 
field in low latitudes,
as seen in the surface radial field of \Fig{fig:testidea}(a).
This oppositely generated polar field reduces the original polar field considerably 
and causes fluctuations in the mean polar field; see \Fig{fig:testidea}(b). 
We note that the polar field shows two spikes after the sudden reduction. This is due
to the fact that the opposite polar field that is produced at low latitudes 
(due to the reversed $\alpha$)
took some time to be transported to the high latitudes, and by that time, 
the polar field was still trying to grow rapidly.  
Anyhow, the abrupt fluctuations in the polar field cause
a reduction in the resulting toroidal field of the next cycle (as the poloidal field 
is the ultimate source of the toroidal field). 
A time delay of about 7 years between the polar field and the toroidal field 
is reflected (\Fig{fig:testidea}) due to the time taken by the meridional flow and diffusion
in transporting the field from the surface to the base of the CZ.
As expected, if the $\alpha_0$ is reversed for a longer time, 
then the double peak becomes more extended; see dotted lines \Fig{fig:testidea}(a) and (b) 
for which the $\alpha$ was reversed for eight months instead of six months.
Interestingly, this simulation spent only two extra months with reversed $\alpha$
but produced much deeper double peaks than the other one. The reason is that model 
spent extra two months near the heightened level of polar field.
In summary, when the polar field 
is growing, a large reduction
in the $\alpha$, causes a double-peak in the next solar cycle.


\subsection{Results of stochastically forced dynamo simulations}
\label{sec:simu}

\begin{figure*}
\centering
\includegraphics[scale=0.75]{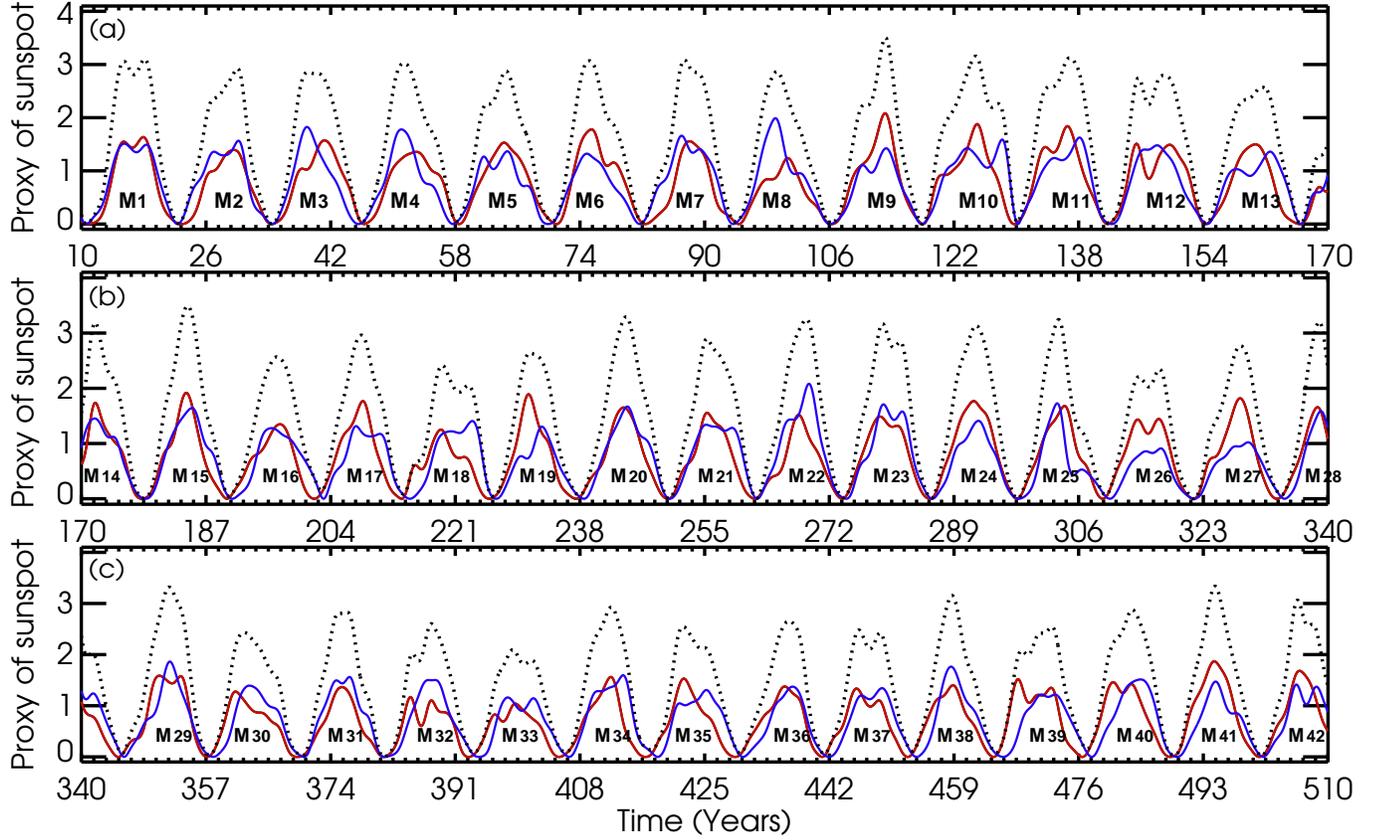}
\caption{
Proxy of sunspot number obtained from our stochastically forced dynamo model.
Dotted (black), red and blue correspond to the total,
northern, and southern hemispheric sunspot numbers, respectively. 
Cycles are numbered with labels M$\#$  to facilitate the discussion.}
\label{fig:modssn}
\end{figure*}

We now model the solar cycle by varying $\alpha$ stochastically in \Eq{eq:pol}.
We replace $\alpha_0$ in \Eq{eq:alpha} by $\alpha_0 + \alpha_\mathrm{fluc} \sigma(t,\tau_\mathrm{corr})$,
where $\alpha_\mathrm{fluc} = $100~\mps, i.e., 200$\%$ level of fluctuations, 
$\sigma$ is a uniform random deviate  whose value lies 
between $-$1 and 1, and $\tau_\mathrm{corr}$ is the correlation time after which the fluctuation is updated to a new value.
As the mean lifetime of the BMRs is about a month, we take $\tau_\mathrm{corr} =$ 1 month. We note
that recently \citet{Kit18} also suggest that to match the distribution of the observed cycle
period, the coherence time of the $\alpha$ fluctuations has to be around 1 solar rotation (25.4 days).
We further note that in this case the fluctuations are introduced independently 
in two hemispheres as the fluctuations in \bl\ process is expected to be uncorrelated in hemispheres.
With this level of fluctuations in $\alpha$, 
our model produces a variation (about $46\%$) in the peak polar field that is comparable 
to the variation ($52\%$) computed from the proxy of the polar field presented in \citet{muno12}.

A result of 500~years simulation of our stochastically forced dynamo model is shown in  
\Fig{fig:modssn}. 
We do not understand how to translate the toroidal field 
in the deep interior to the observed sunspot number.
We use the prescription followed by \citet{CD00}
and build a proxy of the
sunspot number in the northern (southern) hemisphere from the magnetic energy density at
$15^\circ$ ($-15^\circ$) latitude at the base of the convection zone ($r = 0.7R$).

At a first glimpse, we find that the model beautifully reproduces the observed solar cycle with
double peaks in many solar cycles. Not only double peaks, some cycles, in fact, show multiple peaks
around the solar maxima and spikes in the rising and declining phase of the solar cycle.
As seen in the observed solar cycle data, the double-peak may not necessarily
occur in two hemispheres simultaneously.  Only for four cycles in this figure, 
the double-peak appeared in both hemispheres (cycles: M1, M11, M26, M33, and M37).
For other cycles, the double/multiple peaks appear only in one hemisphere.
As discussed in the Introduction and seen in the observed data, when two hemispheres are not synchronized and two hemispheric maxima have a time difference, we may see a double-peak.
We observe that the maxima of cycles: M3, M4, M8, M10, M11, M13, M18, M19, M22, M30, M32, M35, M39, and M40
have significant time lags,
but only cycles M10, M18, M32, M35, and M39 are clearly double-peaked, while the rest are not.
Thus, merely a phase lag of two hemispheric activities may not necessarily
lead to a double-peaked solar cycle.

\subsection{Are the results sensitive to the details of the model?}
One may wonder whether our modelled double-peaked solar cycles presented in \Sec{sec:res} 
are sensitive to the details of the parameters.
To check this, we perform several simulations at different values of parameters.
First, we do two simulations: in one, we reduce the diffusivity of the poloidal field
to its half i.e., $\eta_0 = 1\times 10^{12}$~\cmss\ and in another, we double the value. 
As the dynamo growth rate is largely dependent on the value of diffusivity, we also need to change
the value of $\alpha_0$ in these simulations to 10~\mps\ and 80~\mps, respectively.
We find that the higher-diffusivity simulation 
produces less prominent and infrequent double peaks, while the lower-diffusivity one produces
very frequent and pronounced double peaks.
This is expected because the diffusion tries to smooth out the fluctuations acquired
in the poloidal field.
However, when the level of fluctuations is increased, frequent double peaks
again appear even in the higher-diffusivity simulation.
Next, we execute the following three simulations:  (i) at two different values of
the speed of meridional flow,
namely $v_0 = 20$~\mps\ and $v_0 = 32$~\mps (instead of 26~\mps\ as used in the main simulation),
and (ii) one at $\alpha_0 = 30$~\mps
(instead of 50~\mps). 
No other parameters are changed in these runs.
All three simulations produced qualitatively similar double peaks and spikes as shown in \Fig{fig:modssn}.
Then we perform four simulations at 
$50\%$, $100\%$, $150\%$, and $250\%$ levels of fluctuations in $\alpha$.
Obviously, when the level of fluctuations is increased,
we get more frequent and prominent double peaks
and vice versa. We find that when the fluctuation level is reduced 
below $100\%$, double peaks and spikes disappear;
see \Fig{fig:diffpara} for the results of simulations 
with $100\%$ and $250\%$ levels of fluctuations in $\alpha$.
Finally, we change the coherence time $\tau_\mathrm{corr}$ and  
perform two simulations at $\tau_\mathrm{corr} = 15$ days and 2 months.
We find that larger $\tau_\mathrm{corr}$ produces more prominent and frequent double peaks
(and vice versa).
We also perform a simulation by including 4~\mps\ downward magnetic 
pumping in our model in the same way as done in
\citet{KC16} and find that our final conclusion remains unchanged. In fact,
we find that the downward magnetic pumping helping to transport the polar flux
efficiently to the base of the CZ and thus helping to produce more prominent double peaks.

\section{Observational supports of the idea}
\label{sec:supp}

\begin{figure}
\centering
\includegraphics[scale=0.55]{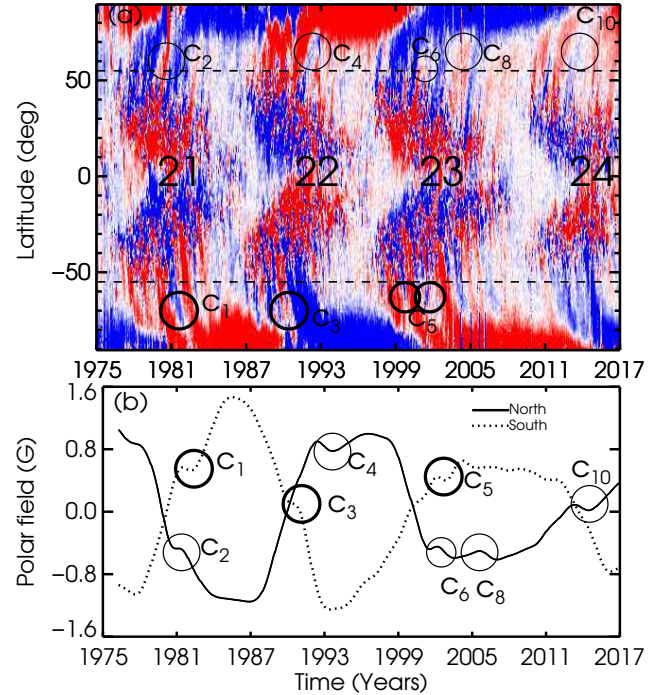}
\caption{
(a) Butterfly diagram of the surface radial magnetic field 
obtained from the National Solar Observatory (NSO/KPVT and SOLIS). 
The blue and red color represent the negative and the positive fields. 
(b) Temporal variation of the Wilcox Solar Observatory (WSO) 
polar field observations for the two hemispheres. 
These polar measurements are constituted from 55$^\circ$ latitude and 
above for both the hemispheres. 
Data source: \url{http://wso.stanford.edu/Polar.html}.
}
\label{fig:obs_pol}
\end{figure}

In previous sections, we have shown that large negative fluctuations in 
the $\alpha$ effect can introduce a large reduction in the poloidal field
which ultimately causes double peaks (including multiple peaks and spikes)
in the solar cycle. Is this really happening in the sun and is there any observational evidence for that? 
This is exactly what we explore here. 
In the \bl\ process, the poloidal field is generated from the decay and dispersal of the tilted BMRs on the solar surface which is seen in the solar observations \citep{Das10,KO11,CS15}. 
The BMR tilt, however, has a large scatter around Joy's law \citep{How91,SK12,MNL14,pavai15,MN16}. 
Due to this scatter, BMRs can occasionally get wrong tilts and 
can produce opposite polarity polar field.
This is actually seen in the observed magnetic field data. 
In \Fig{fig:obs_pol}, circles mark the opposite polarity surges
possibly caused by the wrongly tilted sunspots.
Because of these opposite polarity surges, we see clear fluctuations
in the mean polar field, as shown by corresponding circles in 
\Fig{fig:obs_pol}(b). These fluctuations 
in the polar field are the cause of the observed double peaks in
the subsequent sunspot cycle. 

For example, fluctuations in the polar field in cycle 21, 
as marked by c$_1$ and c$_2$ in the southern and northern hemispheres
(\Fig{fig:obs_pol})
are possibly the cause of the observed double peaks in sunspot cycle~22, 
as shown in \Fig{fig:obsssn}.
Similarly, the fluctuations marked by c$_3$ and c$_4$
in the southern and northern hemispheres are the cause 
of the double peaks in
sunspot cycle 23 in their respective hemispheres as seen in \Fig{fig:obsssn}.
Base on our theoretical experiment presented in \Fig{fig:testidea}, 
we emphasize that a 
large fluctuation in the polar field growing phase is crucial.
Therefore, the wrong polar surge in the southern hemisphere of cycle 22,
as marked by c3 in \Fig{fig:obs_pol}(a) was enough to produce 
a prominent spike in the subsequent sunspot cycle 23 of the same 
hemisphere (\Fig{fig:obsssn}). 

In cycle 23 southern hemisphere polar field, two little negative 
surges marked by C$_5$
produce a little dip in the polar field as shown in \Fig{fig:obs_pol}(b)
which in turn stalled the rising phase of sunspot cycle 24 
of the same hemisphere.
In the northern hemisphere polar field, two surges marked by
C$_6$ and C$_8$, produce two little dips in the mean polar field as shown 
in \Fig{fig:obs_pol}(b). However, the level of the fluctuations in these two cases
are so weak that no detectable double peak is seen in the subsequent sunspot 
cycle~24 (\Fig{fig:obsssn} north).
In this case, the overall polar field was not able to grow due to many negative 
polarity surges and consequently, the solar cycle 24 decreased rapidly 
in the northern hemisphere. 
From \Fig{fig:obs_pol}(b), another fact we discover is that the rapid
build up of the polar field of cycle 22 and 23 
in the northern hemisphere helped to peak
the subsequent sunspot cycles 23 and 24, respectively in the same hemisphere first.

Finally, we notice that there is a prominent surge in the northern hemisphere
of cycle 24 polar field as marked by C$_{10}$ in \Fig{fig:obs_pol}. Therefore,
based on our theoretical model as well as observations, we predict that 
in the northern hemisphere of the forthcoming solar cycle 25 will have a 
dip in the rising phase.

We have seen that our theoretical idea is supported by the available
observed data of the last three cycles. Now the question is, does this idea work also in 
previous cycles when we do not have a direct measurement of the polar field? 
On making a careful inspection of a proxy of the polar field as presented in 
Figure~14 of \citet{muno12} and our sunspot cycles shown in \Fig{fig:obsssn}, 
we find that our idea holds also for the previous cycles.
For example, fluctuations in the proxy of polar field data (Figure 14 of \citet{muno12}) 
around 1910 (both hemispheres), 1920 (north), 1930 (both), 1942 (north), 
1960 (both), and 1976 (south), respectively are possibly responsible for 
the double peak(s)/spike(s) in the corresponding hemisphere
of subsequent sunspot cycles 15, 16, 17, 18, 20, and 21.
In \Fig{fig:obsssn}, we recall that the strongest sunspot cycle 19 did not show any double-peak
or prominent spikes in both the hemispheres. Interestingly, the polar field (around 1955) 
for the cycle 19
also did not show any significant fluctuations. 
In fact, a little halt, in the northern hemisphere polar field around 1950,
is probably caused a little spike in the same hemisphere sunspot number around 1958.
Although promising, we must remember that a detailed comparison with polar field and the subsequent solar cycle
may be misleading as this polar field data is not the actual measurement but a proxy and also it is poorly binned. 

%
\section{Conclusion and Discussion}
\label{sec:conc}

In this study, we have theoretically modelled the double peaks and spikes 
observed in the solar cycle. We have shown that due to large negative fluctuations in the \bl\
process can abruptly decrease or even reverse the polar field for a short time.
This is observed in the form of frequent polar surges of wrong polarity field in surface 
polar field data available for last four solar cycles (\Sec{sec:supp}).
The proxy of polar field data for the previous cycles 
(for which the polar field measurement is not available; \citet{muno12}) 
also shows occasional fluctuations. 
As the polar field is the seed for the next cycle, the fluctuations in the polar field
can be propagated in the subsequent solar cycle and 
they can cause short-term fluctuations in the solar cycle. When the 
abrupt decrease in the polar field happens in the growing phase of the polar field,
we observe a clear double peak in the subsequent sunspot cycle.

We have presented this idea by making a clean experiment
in which we have artificially flipped the source of the poloidal field ($\alpha$)
for six months and as a result a momentary reversed polar field promotes a clear double peak in
the next sunspot cycle. Next, we performed a set of long simulations by including random scatter
in the $\alpha$ and reproduce many double-peaked solar cycles.

To the best of our knowledge, this is the first systematic effort of modeling the                  
double-peaked solar cycle, although three previous attempts exist.
First, \citet{Gnev67} argued that the double-peaked solar cycle is caused by two different
processes occurring at two different latitude bands. When the time interval between
maxima of these two processes is large, the double-peak is seen in the latitude-averaged solar activity.
If this is the correct explanation of double peaks, then the question
remains what are these two physical process and what determines the time lag between them.
Second, \citet{Geor} based on the mechanism of flux transport dynamo showed  
that the double peaks are the manifestation of two surges of the toroidal field.
One surge is generated from the poloidal field that is advected
due to meridional circulation
all the way on the poles, down to the base of CZ and finally to low latitudes,
and the other surge is generated from the poloidal field that is
diffused to the base of the CZ directly from the surface. She suggested that
when the timescales involved in two surges of toroidal field
do not coincide, double peak in the sunspot cycle is observed.
However, no modelled double-peak solar cycle was presented.
To our knowledge, if this idea applies to the flux transport dynamo model,
then we would have observed double peak even without including fluctuations 
in the \bl\ process. Without introducing fluctuations in $\alpha$, we however
have not observed any double peak in any simulations in the parameter ranges 
we have explored.
Therefore, this idea does not work, at least, in our model.
Finally, quasi-periodic nonlinear oscillations in the tachocline \citep{Dik17,Dik18}, as discussed in
the Introduction, could be a possible cause of the doubled peaked solar cycle,
although a detailed model is needed.

One strong support of our idea is that the fluctuations in the \bl\ 
process are identified in the observed polar field as well as in the proxy 
of the polar field data (as discussed in \Sec{sec:supp}). 
Recent independent studies \citep{Ca13, JCS15, Mord16, Kit18} also support the polar field 
fluctuations and they have proposed that these 
fluctuations are the cause of the variation in the subsequent solar cycle. 
Another strong support is the fact that the double peaks are observed independently in two hemispheres and in any phase of the solar cycle. 
If the double peaks are caused by the fluctuations in the dynamo process, then 
they are expected to appear in any hemisphere and in fact, in any phase of the solar cycle. 
This is exactly observed in the Sun.  Occasional
spikes or dips observed in the rising or declining phases of the solar cycle
are also caused by the same origin.

\begin{appendix}
\begin{figure}
\centering
\includegraphics[scale=0.75]{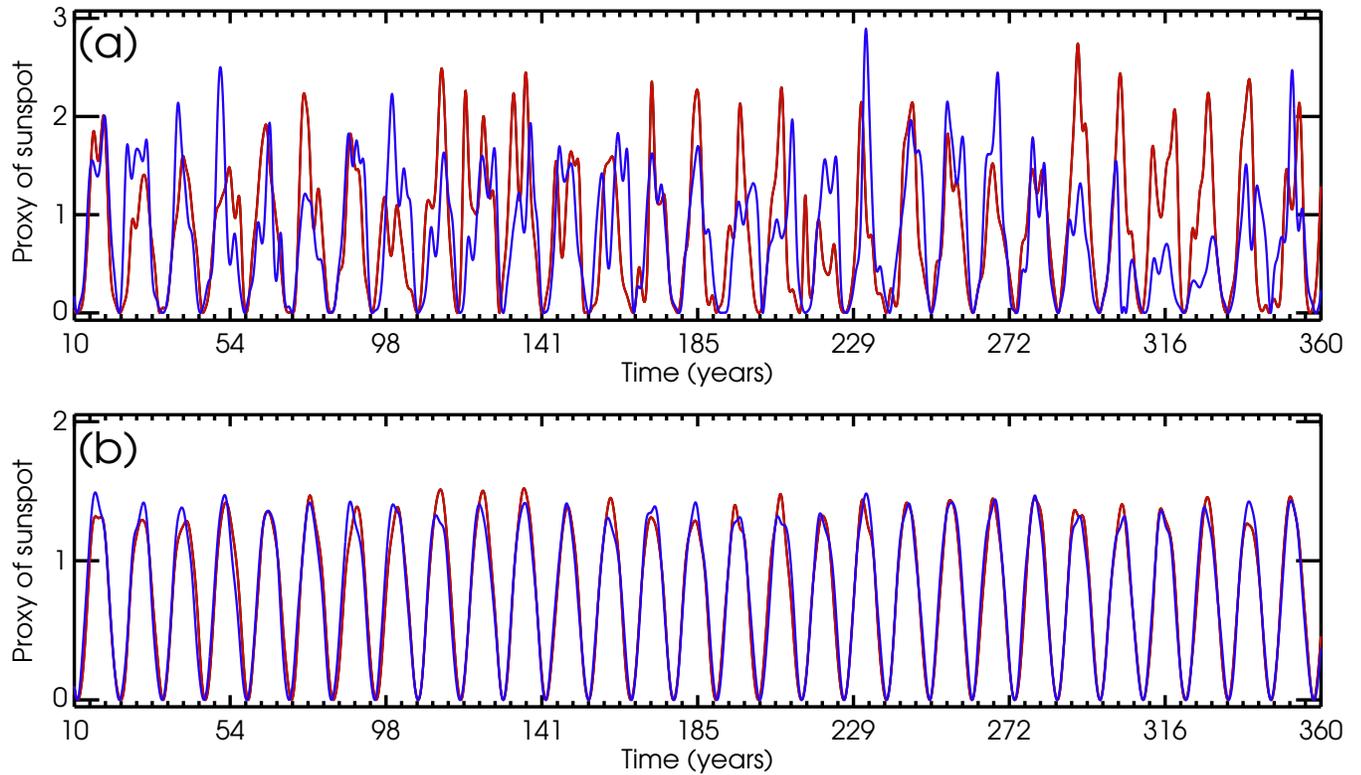}
\caption{
(a) Same as Figure 3 but obtained from simulations with $250\%$ (top) and $100\%$ 
fluctuations in $\alpha$.
}
\label{fig:diffpara}
\end{figure}


\end{appendix}

\begin{acknowledgements}
It is our pleasure to thank Tanmoy Samanta and Gopal Hazra who encouraged us to work on this project.
We are thankful to Sanjay Gosain of NSO for providing the magnetic field data
utilized in making \Fig{fig:obs_pol}(a). 
We also thank anonymous referee for providing critical comments which helped to clarify the presentation of the paper.
BBK and DB acknowledge the research grant: Indo-US (IUSSTF) Joint Networked R\&D Center (Ref: IUSSTF-JC-011-2016).
\end{acknowledgements}

\bibliography{paper}

\begin{thebibliography}{}
\expandafter\ifx\csname natexlab\endcsname\relax\def\natexlab#1{#1}\fi

\bibitem[{{Bazilevskaya} {et~al.}(2014){Bazilevskaya}, {Broomhall}, {Elsworth},
  \& {Nakariakov}}]{Baz14}
{Bazilevskaya}, G., {Broomhall}, A.-M., {Elsworth}, Y., \& {Nakariakov}, V.~M.
  2014, \ssr, 186, 359

\bibitem[{{Cameron} \& {Sch{\"u}ssler}(2015)}]{CS15}
{Cameron}, R., \& {Sch{\"u}ssler}, M. 2015, Science, 347, 1333

\bibitem[{{Cameron} {et~al.}(2013){Cameron}, {Dasi-Espuig}, {Jiang}, {I{\c
  s}{\i}k}, {Schmitt}, \& {Sch{\"u}ssler}}]{Ca13}
{Cameron}, R.~H., {Dasi-Espuig}, M., {Jiang}, J., {et~al.} 2013, \aap, 557,
  A141

\bibitem[{{Charbonneau}(2010)}]{Cha10}
{Charbonneau}, P. 2010, Liv. Rev. Sol. Phys., 7, 3

\bibitem[{{Charbonneau} \& {Dikpati}(2000)}]{CD00}
{Charbonneau}, P., \& {Dikpati}, M. 2000, \apj, 543, 1027

\bibitem[{{Chatterjee} {et~al.}(2004){Chatterjee}, {Nandy}, \&
  {Choudhuri}}]{CNC04}
{Chatterjee}, P., {Nandy}, D., \& {Choudhuri}, A.~R. 2004, \aap, 427, 1019

\bibitem[{{Choudhuri}(2003)}]{Ch03}
{Choudhuri}, A.~R. 2003, \solphys, 215, 31

\bibitem[{{Choudhuri} {et~al.}(2007){Choudhuri}, {Chatterjee}, \&
  {Jiang}}]{CCJ07}
{Choudhuri}, A.~R., {Chatterjee}, P., \& {Jiang}, J. 2007, Physical Review
  Letters, 98, 131103

\bibitem[{{Choudhuri} \& {Karak}(2009)}]{CK09}
{Choudhuri}, A.~R., \& {Karak}, B.~B. 2009, Res. Astron. Astrophys., 9, 953

\bibitem[{{Choudhuri} {et~al.}(1995){Choudhuri}, {Sch\"ussler}, \&
  {Dikpati}}]{CSD95}
{Choudhuri}, A.~R., {Sch\"ussler}, M., \& {Dikpati}, M. 1995, \aap, 303, L29

\bibitem[{{Dasi-Espuig} {et~al.}(2010){Dasi-Espuig}, {Solanki}, {Krivova},
  {Cameron}, \& {Pe{\~n}uela}}]{Das10}
{Dasi-Espuig}, M., {Solanki}, S.~K., {Krivova}, N.~A., {Cameron}, R., \&
  {Pe{\~n}uela}, T. 2010, \aap, 518, A7

\bibitem[{{Dikpati} {et~al.}(2017){Dikpati}, {Cally}, {McIntosh}, \&
  {Heifetz}}]{Dik17}
{Dikpati}, M., {Cally}, P.~S., {McIntosh}, S.~W., \& {Heifetz}, E. 2017,
  Scientific Reports, 7, 14750

\bibitem[{{Dikpati} {et~al.}(2002){Dikpati}, {Corbard}, {Thompson}, \&
  {Gilman}}]{Dik02}
{Dikpati}, M., {Corbard}, T., {Thompson}, M.~J., \& {Gilman}, P.~A. 2002,
  \apjl, 575, L41

\bibitem[{{Dikpati} \& {Gilman}(2001)}]{DG01}
{Dikpati}, M., \& {Gilman}, P.~A. 2001, \apj, 559, 428

\bibitem[{{Dikpati} \& {Gilman}(2009)}]{DG09}
---. 2009, \ssr, 144, 67

\bibitem[{{Dikpati} {et~al.}(2018){Dikpati}, {McIntosh}, {Bothun}, {Cally},
  {Ghosh}, {Gilman}, \& {Umurhan}}]{Dik18}
{Dikpati}, M., {McIntosh}, S.~W., {Bothun}, G., {et~al.} 2018, \apj, 853, 144

\bibitem[{{D'Silva} \& {Choudhuri}(1993)}]{DC93}
{D'Silva}, S., \& {Choudhuri}, A.~R. 1993, \aap, 272, 621

\bibitem[{{Durney}(1995)}]{Dur95}
{Durney}, B.~R. 1995, \solphys, 160, 213

\bibitem[{{Feminella} \& {Storini}(1997)}]{FS97}
{Feminella}, F., \& {Storini}, M. 1997, \aap, 322, 311

\bibitem[{{Georgieva}(2011)}]{Geor}
{Georgieva}, K. 2011, ISRN Astronomy and Astrophysics, 2011, 437838

\bibitem[{{Gnevyshev}(1963)}]{Gnev63}
{Gnevyshev}, M.~N. 1963, \sovast, 7, 311

\bibitem[{{Gnevyshev}(1967)}]{Gnev67}
---. 1967, \solphys, 1, 107

\bibitem[{{Gnevyshev}(1977)}]{Gnev77}
---. 1977, \solphys, 51, 175

\bibitem[{{Hazra} {et~al.}(2017){Hazra}, {Choudhuri}, \& {Miesch}}]{HCM17}
{Hazra}, G., {Choudhuri}, A.~R., \& {Miesch}, M.~S. 2017, \apj, 835, 39

\bibitem[{{Howard}(1991)}]{How91}
{Howard}, R.~F. 1991, \solphys, 136, 251

\bibitem[{{Jiang} {et~al.}(2014){Jiang}, {Cameron}, \& {Sch{\"u}ssler}}]{JCS14}
{Jiang}, J., {Cameron}, R.~H., \& {Sch{\"u}ssler}, M. 2014, \apj, 791, 5

\bibitem[{{Jiang} {et~al.}(2015){Jiang}, {Cameron}, \& {Sch{\"u}ssler}}]{JCS15}
---. 2015, \apjl, 808, L28

\bibitem[{{Kane}(2009)}]{Kane09}
{Kane}, R.~P. 2009, Annales Geophysicae, 27, 1469

\bibitem[{{Kane}(2010)}]{Kane10}
---. 2010, \solphys, 261, 209

\bibitem[{{Karak} \& {Cameron}(2016)}]{KC16}
{Karak}, B.~B., \& {Cameron}, R. 2016, \apj, 832, 94

\bibitem[{{Karak} {et~al.}(2014){Karak}, {Jiang}, {Miesch}, {Charbonneau}, \&
  {Choudhuri}}]{Kar14a}
{Karak}, B.~B., {Jiang}, J., {Miesch}, M.~S., {Charbonneau}, P., \&
  {Choudhuri}, A.~R. 2014, \ssr, 186, 561

\bibitem[{{Karak} \& {Miesch}(2017)}]{KM17}
{Karak}, B.~B., \& {Miesch}, M. 2017, \apj, 847, 69

\bibitem[{{Karak} \& {Miesch}(2018)}]{KM18}
---. 2018, \apjl, 860, L26

\bibitem[{{Karak} \& {Nandy}(2012)}]{KN12}
{Karak}, B.~B., \& {Nandy}, D. 2012, \apjl, 761, L13

\bibitem[{{Kitchatinov} {et~al.}(2018){Kitchatinov}, {Mordvinov}, \&
  {Nepomnyashchikh}}]{Kit18}
{Kitchatinov}, L.~L., {Mordvinov}, A.~V., \& {Nepomnyashchikh}, A.~A. 2018,
  \aap, 615, A38

\bibitem[{{Kitchatinov} \& {Nepomnyashchikh}(2016)}]{KN16}
{Kitchatinov}, L.~L., \& {Nepomnyashchikh}, A.~A. 2016, Advances in Space
  Research, 58, 1554

\bibitem[{{Kitchatinov} \& {Olemskoy}(2011)}]{KO11}
{Kitchatinov}, L.~L., \& {Olemskoy}, S.~V. 2011, Astronomy Letters, 37, 656

\bibitem[{{Lean} \& {Brueckner}(1989)}]{LB89}
{Lean}, J.~L., \& {Brueckner}, G.~E. 1989, \apj, 337, 568

\bibitem[{{Lemerle} \& {Charbonneau}(2017)}]{LC17}
{Lemerle}, A., \& {Charbonneau}, P. 2017, \apj, 834, 133

\bibitem[{{Mandal} {et~al.}(2017){Mandal}, {Hegde}, {Samanta}, {Hazra},
  {Banerjee}, \& {Ravindra}}]{Maetal}
{Mandal}, S., {Hegde}, M., {Samanta}, T., {et~al.} 2017, \aap, 601, A106

\bibitem[{{McClintock} \& {Norton}(2016)}]{MN16}
{McClintock}, B.~H., \& {Norton}, A.~A. 2016, \apj, 818, 7

\bibitem[{{McClintock} {et~al.}(2014){McClintock}, {Norton}, \& {Li}}]{MNL14}
{McClintock}, B.~H., {Norton}, A.~A., \& {Li}, J. 2014, \apj, 797, 130

\bibitem[{{McIntosh} {et~al.}(2015){McIntosh}, {Leamon}, {Krista}, {Title},
  {Hudson}, {Riley}, {Harder}, {Kopp}, {Snow}, {Woods}, {Kasper}, {Stevens}, \&
  {Ulrich}}]{Scott15}
{McIntosh}, S.~W., {Leamon}, R.~J., {Krista}, L.~D., {et~al.} 2015, Nature
  Communications, 6, 6491

\bibitem[{{Miesch} \& {Dikpati}(2014)}]{MD14}
{Miesch}, M.~S., \& {Dikpati}, M. 2014, \apjl, 785, L8

\bibitem[{{Mordvinov} {et~al.}(2016){Mordvinov}, {Pevtsov}, {Bertello}, \&
  {Petri}}]{Mord16}
{Mordvinov}, A., {Pevtsov}, A., {Bertello}, L., \& {Petri}, G. 2016,
  Solar-Terrestrial Physics, 2, 3

\bibitem[{{Mu{\~n}oz-Jaramillo} {et~al.}(2013){Mu{\~n}oz-Jaramillo},
  {Dasi-Espuig}, {Balmaceda}, \& {DeLuca}}]{Muno13}
{Mu{\~n}oz-Jaramillo}, A., {Dasi-Espuig}, M., {Balmaceda}, L.~A., \& {DeLuca},
  E.~E. 2013, \apjl, 767, L25

\bibitem[{{Mu{\~n}oz-Jaramillo} {et~al.}(2012){Mu{\~n}oz-Jaramillo}, {Sheeley},
  {Zhang}, \& {DeLuca}}]{muno12}
{Mu{\~n}oz-Jaramillo}, A., {Sheeley}, N.~R., {Zhang}, J., \& {DeLuca}, E.~E.
  2012, \apj, 753, 146

\bibitem[{{Nagy} {et~al.}(2017){Nagy}, {Lemerle}, {Labonville}, {Petrovay}, \&
  {Charbonneau}}]{Nagy17}
{Nagy}, M., {Lemerle}, A., {Labonville}, F., {Petrovay}, K., \& {Charbonneau},
  P. 2017, \solphys, 292, 167

\bibitem[{{Norton} \& {Gallagher}(2010)}]{NG10}
{Norton}, A.~A., \& {Gallagher}, J.~C. 2010, \solphys, 261, 193

\bibitem[{{Priyal} {et~al.}(2014){Priyal}, {Banerjee}, {Karak},
  {Mu{\~n}oz-Jaramillo}, {Ravindra}, {Choudhuri}, \& {Singh}}]{Priy14}
{Priyal}, M., {Banerjee}, D., {Karak}, B.~B., {et~al.} 2014, \apjl, 793, L4

\bibitem[{{Senthamizh Pavai} {et~al.}(2015){Senthamizh Pavai}, {Arlt},
  {Dasi-Espuig}, {Krivova}, \& {Solanki}}]{pavai15}
{Senthamizh Pavai}, V., {Arlt}, R., {Dasi-Espuig}, M., {Krivova}, N.~A., \&
  {Solanki}, S.~K. 2015, \aap, 584, A73

\bibitem[{{Stenflo} \& {Kosovichev}(2012)}]{SK12}
{Stenflo}, J.~O., \& {Kosovichev}, A.~G. 2012, \apj, 745, 129

\bibitem[{{Yeates} {et~al.}(2008){Yeates}, {Nandy}, \& {Mackay}}]{YNM08}
{Yeates}, A.~R., {Nandy}, D., \& {Mackay}, D.~H. 2008, \apj, 673, 544

\bibitem[{{Zaqarashvili}(2018)}]{Za18}
{Zaqarashvili}, T. 2018, \apj, 856, 32

\bibitem[{{Zaqarashvili} {et~al.}(2010){Zaqarashvili}, {Carbonell}, {Oliver},
  \& {Ballester}}]{Za10}
{Zaqarashvili}, T.~V., {Carbonell}, M., {Oliver}, R., \& {Ballester}, J.~L.
  2010, \apj, 709, 749

\end{thebibliography}
\end{document}